\documentstyle[preprint,aps,epsf,floats]{revtex}

\begin{document}
\tighten

\newcommand{\gsim}{\raisebox{-0.7ex}{$\stackrel{\textstyle >}{\sim}$ }}
\newcommand{\lsim}{\raisebox{-0.7ex}{$\stackrel{\textstyle <}{\sim}$ }}

\def\Journal#1#2#3#4{{#1} {\bf #2}, #3 (#4)}

\def\NCA{\em Nuovo Cimento}
\def\NIM{\em Nucl. Instrum. Methods}
\def\NIMA{{\em Nucl. Instrum. Methods} A}
\def\NPB{{\em Nucl. Phys.} B}
\def\NPA{{\em Nucl. Phys.} A}
\def\NP{{\em Nucl. Phys.} }
\def\PLB{{\em Phys. Lett.} B}
\def\PRL{\em Phys. Rev. Lett.}
\def\PRD{{\em Phys. Rev.} D}
\def\PRC{{\em Phys. Rev.} C}
\def\PRA{{\em Phys. Rev.} A}
\def\PR{{\em Phys. Rev.} }
\def\ZPC{{\em Z. Phys.} C}
\def\SJP{{\em Sov. Phys. JETP}}
\def\SJNP{{\em Sov. Phys. Nucl. Phys.}}

\def\FBS{{\em Few Body Systems Suppl.}}
\def\IJMP{{\em Int. J. Mod. Phys.} A}
\def\UJP{{\em Ukr. J. of Phys.}}
\def\CJP{{\em Can. J. Phys.}}
\def\SCI{{\em Science} }
\def\AST{{\em Astrophys. Jour.} }

\preprint{\vbox{
\hbox{ NT@UW-01-013}
}}
\bigskip
\bigskip

\title{Chiral Corrections to Matrix Elements of Twist-2 Operators}
\author{{\bf Daniel Arndt}$^a$ and {\bf Martin J. Savage}$^{a,b}$}
\address{$^a$ Department of Physics, University of Washington, \\
Seattle, WA 98195. }
\address{$^b$ Jefferson Laboratory, 12000 Jefferson Avenue,\\
Newport News, VA 23606.}
\maketitle

\begin{abstract}
We compute the leading non-analytic contributions of the form 
$m_q\log m_q$ to matrix elements of twist-2 
operators in the nucleon and pion using effective field theory.
Previously omitted one-loop contributions that are 
related to tree-level matrix elements by chiral symmetry
are included.
\end{abstract}

\vskip 2in

\vfill\eject

\section{Introduction}

Deep-Inelastic-Scattering (DIS) from nucleon targets has provided  
a wealth of
information about the nature  of 
strong interactions and the structure of the nucleon.
From the initial  discovery of partons in the 1970's, 
DIS continues to provide ever more precise 
measurements of the parton-distribution functions (PDFs).
It is highly desirable to make  a direct connection between the
experimental data and the 
now well-established theory of strong interactions, QCD.
A rigorous and model-independent comparison will
be accomplished by 
performing high-statistics unquenched or partially-quenched~\cite{Pqqcd}
lattice-QCD calculations~\cite{latticeQCD}.
Presently, lattice calculations
cannot be performed with the physical values of
the light quark masses, $m_q$, 
($m_u\sim 5~{\rm MeV}$, $m_d\sim 10~{\rm MeV}$)
and extrapolations from the lattice masses,
that produce a pion of mass $m_\pi^{\rm latt.}\sim 500~{\rm MeV}$,
to the physical values must be performed.
Of course, 
such extrapolations require knowledge of the $m_q$-dependence of the
matrix element of interest. 
Recently hadronic models, such as the Cloudy Bag
Model (CBM), have been used to motivate
explicit forms for the $m_q$-dependence of forward 
matrix elements of the non-singlet twist-2 operators
that  contribute to DIS from the nucleon~\cite{CBMa,CBM}.
In addition, these models have been used to connect lattice calculations to
other properties of the nucleon, such as electromagnetic form 
factors~\cite{HJLT}.

It is well established that one can determine the $m_q$-dependence of hadronic 
observables by performing a systematic expansion about the 
chiral limit~\cite{chiralpi,JMheavy,chiralN,chiralUlf}.
In fact, extensive work has been accomplished in understanding the properties
and interactions of the low-lying mesons and baryons in both two-flavour and
three-flavour QCD,
such as the magnetic moments~\cite{chiralMAG},
the electric form factors~\cite{chiralEff},
the axial matrix elements~\cite{chiralAX},
and the polarizabilities of the nucleons and other 
octet baryons~\cite{chiralPOL},
to name just a few,
and the analogous quantities in hadrons containing heavy quarks~\cite{chiralQ}.
In addition, there has been an extensive effort during the last decade to 
include multi-nucleon systems in this framework~\cite{multiN}.
In this work, we include
twist-2 operators in the chiral lagrangian, 
and compute the leading non-analytic contributions of the 
form $m_q\log m_q$ to their matrix elements.

The pion fields are introduced 
into the low-energy effective field theory (EFT),
chiral perturbation theory ($\chi$PT),
through 
the $\Sigma$-field,
\begin{eqnarray}
\Sigma & = & 
\exp\left({2 i M\over f}\right)
\ \ \ \ ,\ \ \ \ 
M = \left(\matrix{\pi^0/\sqrt{2} & \pi^+\cr
\pi^- & -\pi^0/\sqrt{2}}\right)
\ =\ {1\over \sqrt{2}} \ \tau^\alpha \ \pi^\alpha
\ \ \ .
\end{eqnarray}
with $f=132~{\rm MeV}$.
$\Sigma=\xi^2$ transforms as 
$\Sigma\rightarrow L\Sigma R^\dagger\ =\ L\xi U^\dagger\ U\xi R^\dagger$
under $SU(2)_L\otimes SU(2)_R$ chiral transformations.
In order to construct an EFT with well-defined power-counting, the
nucleons are treated as heavy fields
in the Heavy-Baryon formulation of Jenkins and Manohar~\cite{JMheavy},
and transform as $N_v\rightarrow U N_v$ (the subscript $v$ denotes the
four-velocity of the nucleon) under chiral transformations (for reviews see
Ref.~\cite{chiralN,chiralUlf}).
Below the chiral symmetry breaking scale $\Lambda_\chi$,
S-matrix elements can be expanded 
in derivatives and in $m_q$.
The naive size of the matrix element of an operator with 
$n_1$ creation operators for heavy nucleons,
$n_2$ annihilation operators for heavy nucleons,
$n_3$ derivatives,
$n_4$ light quark mass matrices,
$n_5$ powers of the nucleon four-velocity $v$,
$n_6$ $\Sigma$-field operators and 
$n_7$ $\Sigma^\dagger$-field operators,
is
\begin{eqnarray}
f^2 \Lambda_\chi^2 
{\left(\bar N_v\over f\sqrt{\Lambda_\chi} \right)}^{n_1}
{\left( N_v\over f\sqrt{\Lambda_\chi}    \right)}^{n_2} 
{\left(\partial\over  \Lambda_\chi \right)}^{n_3}
{\left(m_q\over \Lambda_\chi  \right)}^{n_4}
{\left( v\right)}^{n_5}
{\left( \Sigma\right)}^{n_6}
{\left( \Sigma^\dagger\right)}^{n_7} 
\ \ \ ,
\label{eq:pcrules}
\end{eqnarray}
which should be considered merely as a guide.
The strong interactions between pions and nucleons at leading 
order in the chiral expansion arise from a lagrange density of the 
form
\begin{eqnarray}
{\cal L} & = & 
{f^2\over 8}
{\rm Tr}\left[\ \partial^\mu\Sigma\ \partial_\mu\Sigma^\dagger\ \right]
\ +\ \lambda\  {\rm Tr}\left[\ m_q\Sigma^\dagger \ +\ {\rm h.c.} \ \right]
\nonumber\\
& + & 
\overline{N}_v\ iv\cdot D N_v
\ +\ 
2 g_A \overline{N}_v\ S\cdot {\cal A}\ N_v
\ \ \ ,
\label{eq:strong}
\end{eqnarray}
where $D_\mu = \partial_\mu + {\cal V}_\mu$ 
is the chiral-covariant derivative with
${\cal V}_\mu = {1\over 2}\left(\xi\partial_\mu\xi^\dagger  
+ \xi^\dagger\partial_\mu\xi\right)$
the pion vector field, $g_A=1.25$ is the axial-vector coupling constant, 
$S^\mu$ is the covariant spin-operator defined in Ref.~\cite{JMheavy},
${\cal A}_\mu =  {i\over 2}\left(\xi\partial_\mu\xi^\dagger  
- \xi^\dagger\partial_\mu\xi\right)$ is the axial-vector pion field,
and $\lambda$ is a parameter that provides the leading order contribution
to the pion mass.

\section{Non-Singlet, Twist-2 Operators}
In this section
we will focus on forward matrix elements of the non-singlet twist-2
operators
\begin{eqnarray}
{\cal O}^{(n), a}_{\mu_1\mu_2\ ... \mu_n}
& = & 
{1\over n!}\ 
\overline{q}\ \tau^a\ \gamma_{ \{\mu_1  } 
\left(i \stackrel{\leftrightarrow}{D}_{\mu_2}\right)\ 
... 
\left(i \stackrel{\leftrightarrow}{D}_{ \mu_n\} }\right)\ q
\ -\ {\rm traces}
\ \ \ ,
\label{eq:twistop}
\end{eqnarray}
where the $\{ ... \}$ denotes symmetrization on all Lorentz indices.
Once the transformation properties 
of ${\cal O}^{(n), a}_{\mu_1\mu_2\ ... \mu_n}$
under  $SU(2)_L\otimes SU(2)_R$ have been established,
the operators in the pion-nucleon EFT that reproduce matrix elements 
of ${\cal O}^{(n), a}_{\mu_1\mu_2\ ... \mu_n}$ can be constructed, and further,
the EFT can be used to perform a model-independent calculation of
the low-momentum contributions to its matrix element.
In order to determine its transformation properties it 
is convenient to write it in terms of left-handed and 
right-handed quark fields,
\begin{eqnarray}
{\cal O}^{(n), a}_{\mu_1\mu_2\ ... \mu_n}
& = & 
{\cal O}^{(n), a}_{L,\ \mu_1\mu_2\ ... \mu_n}
\ +\
{\cal O}^{(n), a}_{R,\ \mu_1\mu_2\ ... \mu_n}
\end{eqnarray}
with
\begin{eqnarray}
{\cal O}^{(n), a}_{L,\ \mu_1\mu_2\ ... \mu_n}
& = & 
{1\over n!}\ 
\overline{q}_L\ \tau^a_L\ \gamma_{ \{\mu_1  } 
\left(i \stackrel{\leftrightarrow}{D}_{\mu_2}\right)\ 
... 
\left(i \stackrel{\leftrightarrow}{D}_{ \mu_n\} }\right)\ q_L
\ -\ {\rm traces}
\nonumber\\
{\cal O}^{(n), a}_{R,\ \mu_1\mu_2\ ... \mu_n}
& = & 
{1\over n!}\ 
\overline{q}_R\ \tau^a_R\ \gamma_{ \{\mu_1  } 
\left(i \stackrel{\leftrightarrow}{D}_{\mu_2}\right)\ 
... 
\left(i \stackrel{\leftrightarrow}{D}_{ \mu_n\} }\right)\ q_R
\ -\ {\rm traces}
\ \ \ ,
\label{eq:twistchiral}
\end{eqnarray}
from which it is clear that 
${\cal O}^{(n), a}_{\mu_1\mu_2\ ... \mu_n}$ transforms as 
$(3,1)\oplus (1,3)$ under $SU(2)_L\otimes SU(2)_R$.
For bookkeeping purposes we have introduced the flavour matrices 
$\tau^a_L$ and $\tau^a_R$ that are taken to 
transform as 
$\tau^a_L  \rightarrow  L \tau^a_L L^\dagger$
and 
$\tau^a_R  \rightarrow  R \tau^a_R R^\dagger$
under $SU(2)_L\otimes SU(2)_R$.

\begin{figure}[!ht]
\centerline{{\epsfxsize=3.0in \epsfbox{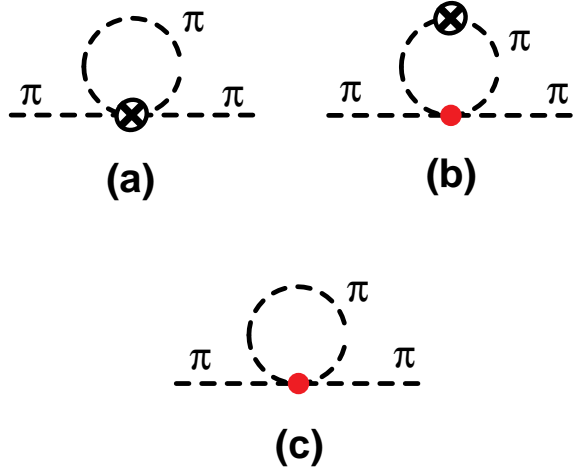}}} 
\vskip 0.15in
\noindent
\caption{\it 
The pion loop diagrams that give the leading non-analytic contributions
to the matrix element of ${\cal O}^{(n), a}_{\mu_1\mu_2\ ... \mu_n}$
between single-pion states.
The crossed circle denotes an insertion of an operator from
eq.~(\ref{eq:meson}), arising directly from the twist-2 operator.
The smaller solid circle denotes an insertion of a leading order 
strong-interaction vertex from eq.~(\ref{eq:strong}).
Diagrams (a) and (b) are vertex corrections while diagram (c) denotes
wavefunction renormalization.
}
\label{fig:pidiags}
\vskip .2in
\end{figure}

If we are interested in DIS from pions, or
in pion loop contributions 
to DIS from the nucleon we require the matrix
element of 
${\cal O}^{(n), a}_{\mu_1\mu_2\ ... \mu_n}$
in the pion.
The power-counting rules of eq.~(\ref{eq:pcrules}) dictate that such matrix
elements will be dominated by operators involving the least number of
derivatives and insertions of $m_q$, but can have an
arbitrary number of insertions of the 
$\Sigma$ and $\Sigma^\dagger$ fields.
Operators that can contribute are of the form
\begin{eqnarray}
{\cal O}^{(n), a}_{\mu_1\mu_2\ ... \mu_n}
&\rightarrow &
a^{(n)} \left(i\right)^n {f^2\over 4}
\left({1\over\Lambda_\chi}\right)^{n-1}
\ {\rm Tr}\left[
\Sigma^\dagger \tau^a \overrightarrow\partial_{\mu_1}
...\overrightarrow\partial_{\mu_n}
\Sigma
\ +\ 
\Sigma 
\tau^a \overrightarrow\partial_{\mu_1}
...\overrightarrow\partial_{\mu_n}
\Sigma^\dagger
\right]
\ -\ {\rm traces}
\nonumber\\
& = & 
a^{(n)} 2 \left(i\right)^n
\left({1\over\Lambda_\chi}\right)^{n-1}
\ i\varepsilon^{\alpha a\beta} \ 
\pi^\alpha  
\overrightarrow\partial_{\mu_1}
...\overrightarrow\partial_{\mu_n} \pi^\beta
\ -\ {\rm traces}
\ +\ {\cal O}\left(\pi^4\right)
\ \ \ .
\label{eq:meson}
\end{eqnarray}
With the exception of $n=1$,
the coefficients $a^{(n)}$ are unknown and 
must be determined elsewhere.
${\cal O}^{(1)}_{\mu_1}$ is the isovector charge operator from which we
deduce that $a^{(1)}=+1$.
In addition to the operators of eq.~(\ref{eq:meson}), there are also operators
of the form
\begin{eqnarray}
{\cal O}^{(n), a}_{\mu_1\mu_2\ ... \mu_n}
\rightarrow 
a^{(n)} \left(i\right)^n {f^2\over 4}
\left({1\over\Lambda_\chi}\right)^{n-1}
& & \ {\rm Tr}\left[
\Sigma^\dagger \tau^a \overrightarrow\partial_{\{ \mu_1}
...\overrightarrow\partial_{\mu_k}
\Sigma
\overrightarrow\partial_{\mu_k}
...\overrightarrow\partial_{\mu_p}
\Sigma^\dagger
.....
\overrightarrow\partial_{\mu_r}
...\overrightarrow\partial_{\mu_n\}}
\Sigma
\right.\nonumber\\
& & \left.
\ +\ 
\Sigma\leftrightarrow\Sigma^\dagger
\right]
\ -\ {\rm traces}
\ \ \ ,
\end{eqnarray}
that must be considered.
However, 
these operators do not contribute to single-pion forward matrix elements
at tree-level,
and only give interactions between three or more pions.
In addition, the symmetry of the Lorentz indices and the tracelessness of 
${\cal O}^{(n), a}_{\mu_1\mu_2\ ... \mu_n}$
means that these operators do not contribute 
to single-pion forward matrix elements
even at one-loop level.
Therefore the diagrams shown in fig.~\ref{fig:pidiags}
give the leading non-analytic contributions to the matrix element 
of ${\cal O}^{(n), a}_{\mu_1\mu_2\ ... \mu_n}$ 
between single-pion states.
After a straightforward calculation, 
one finds that the forward matrix element 
of  ${\cal O}^{(n), a}_{\mu_1\mu_2\ ... \mu_n}$ 
between an initial pion with isospin index $\alpha$ and momentum $q_\mu$, 
and a final pion with isospin index $\beta$ 
vanishes for n-even, and is
\begin{eqnarray}
{\cal M} & = & 
i \ 4\  a^{(n)} 
\ \left({1\over\Lambda_\chi}\right)^{n-1}
\ \left[ 1 - {1-\delta^{n 1}\over 8\pi^2 f^2} m_\pi^2\log\left({m_\pi^2\over
      \Lambda_\chi^2}\right)\ 
+\ ...\ 
\right]
\ \varepsilon^{\alpha\beta a} 
\ q_{\mu_1} ... q_{\mu_n}
\ -\ {\rm traces} 
\ \ \ ,
\label{eq:pitotal}
\end{eqnarray}
for n-odd,
where the ellipses denote terms that are analytic in $m_q$,
or are higher order in the  chiral expansion.
The factor of $1-\delta^{n 1}$ 
in the sub-leading contribution ensures that the 
$n=1$ isospin-charge is not renormalized at loop-level.

\begin{figure}[!ht]
\centerline{{\epsfxsize=3.0in \epsfbox{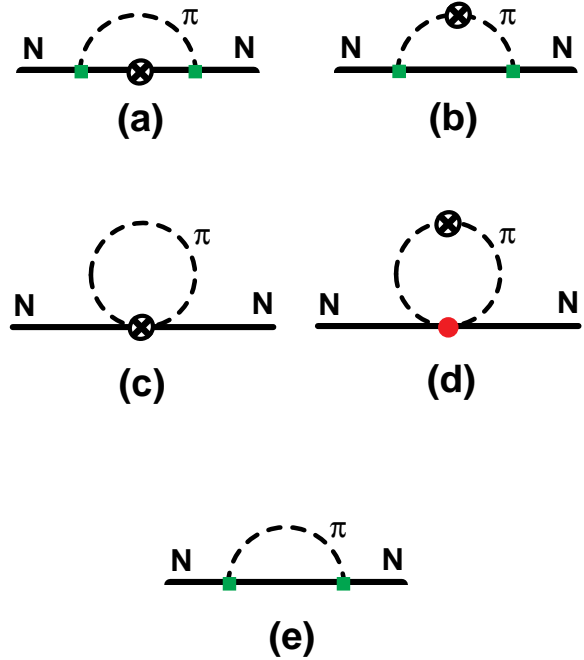}}} 
\vskip 0.15in
\noindent
\caption{\it 
The pion loop diagrams that give the leading non-analytic contributions
to the matrix element of ${\cal O}^{(n), a}_{\mu_1\mu_2\ ... \mu_n}$
between single-nucleon states.
The crossed circle denotes an insertion of an operator from
eq.~(\ref{eq:meson}) or eq.~(\ref{eq:nucleons}), 
arising directly from the twist-2 operator.
The smaller solid circle denotes an insertion of the
strong two-pion-nucleon 
interaction from the nucleon kinetic energy term
in eq.~(\ref{eq:strong}),
while the square denotes an insertion of the axial-vector 
interaction~$\propto g_A$.
Diagrams (a)-(d) are vertex corrections while diagram (e) denotes
nucleon wavefunction renormalization.
}
\label{fig:Ndiags}
\vskip .2in
\end{figure}

The leading order operator contributing to the matrix element of 
${\cal O}^{(n), a}_{\mu_1\mu_2\ ... \mu_n}$
between nucleon states is
\begin{eqnarray}
{\cal O}^{(n), a}_{\mu_1\mu_2\ ... \mu_n}
\ &\rightarrow &\ 
A^{(n)}\ v_{\mu_1} v_{\mu_2} ... v_{\mu_n}\ 
\overline{N}_v \ \tau^a_{\xi+}\ N_v
\ -\ {\rm traces}
\ \ \ ,
\label{eq:nucleons}
\end{eqnarray}
where operators involving more derivatives or insertions of $m_q$ are 
suppressed by powers of $\Lambda_\chi$,
and we have defined $\tau^a_{\xi\pm}$ to be
\begin{eqnarray}
\tau^a_{\xi\pm} & = & 
{1\over 2}\left(\ 
\xi\tau^a\xi^\dagger \pm \xi^\dagger\tau^a\xi\ \right)
\ \ \ .
\label{eq:taudef}
\end{eqnarray}
The coefficients $A^{(n)}$ must be determined elsewhere, except for
$A^{(1)}=+1$ which corresponds to the isospin charge operator.
In addition to the tree-level contribution to the nucleon 
forward matrix element,
there are vertices involving an even number of pion fields.
The diagrams shown in fig.~\ref{fig:Ndiags}
give the leading non-analytic corrections to the forward matrix
element of 
${\cal O}^{(n), a}_{\mu_1\mu_2\ ... \mu_n}$ between single-nucleon states.
A straightforward computation gives
\begin{eqnarray}
{\cal M} & = & 
A^{(n)} \ v_{\mu_1} v_{\mu_2} ... v_{\mu_n}\ 
\overline{U}_v \tau^a U_v\ 
\left[\ 
1\ - \
\left(3 g_A^2 + 1\right)
{1-\delta^{n 1}\over 8\pi^2 f^2} 
 m_\pi^2\log\left({m_\pi^2\over
      \Lambda_\chi^2}\right)\ \right]
\ \ \ ,
\label{eq:Ntotal}
\end{eqnarray}
where $U_v$ is the nucleon spinor, and 
we have only shown the non-analytic part of the sub-leading contribution.
The non-analytic contributions to the $n=1$ matrix element vanish as
the nucleon isospin charge is not renormalized.
Our result in eq.~(\ref{eq:Ntotal}) differs from previous
computations~\cite{CBMa,CBM} primarily due to our inclusion of 
the two-pion-nucleon interaction associated with the 
tree-level matrix element 
of ${\cal O}^{(n), a}_{\mu_1\mu_2\ ... \mu_n}$,
but also in the numerical coefficient of the  $g_A^2$ contribution.
In Ref.~\cite{CBMa} only the contribution from diagram (b) of
fig.~\ref{fig:Ndiags} is computed and correctly found to scale 
as $m_q^{n+1\over 2}\log m_q$.
However, this diagram is only part of the complete result for $n=1$ 
(required by charge conservation) and is sub-dominant for $n>1$.
In Ref.~\cite{CBM}, this was corrected somewhat by the appearance
of $m_q \log m_q$ contributions for all values of $n$, however, the 
$g_A$ independent contributions  were omitted.

It is worth keeping in mind the relative size of non-analytic
terms from loop-diagrams compared with the analytic contributions 
from both loop-diagrams and local counterterms.
The complete set of 
local operators with a single insertion of $m_q$
that contribute to the matrix element of
${\cal O}^{(n), a}_{\mu_1\mu_2\ ... \mu_n}$
is
\begin{eqnarray}
{\cal O}^{(n), a}_{\mu_1\mu_2\ ... \mu_n} & \rightarrow & 
\left(\ 
b_1(\mu)\ \overline{N}\ \{ \tau^a_{\xi +} , \chi_+ \} \ N
\ +\ 
b_2(\mu)\ \overline{N}\ \left[ \tau^a_{\xi +} , \chi_- \right] \ N
\ +\ 
b_3(\mu)\ {\rm Tr}\left[\chi_+\right]\ 
\overline{N}\ \tau^a_{\xi +}\ N
\right.
\nonumber\\
& & + 
\left.
b_4(\mu)\ \overline{N}\ \{ \tau^a_{\xi -} , \chi_- \} \ N
\ +\ 
b_5(\mu)\ \overline{N}\ \left[ \tau^a_{\xi -} , \chi_+ \right] \ N
\ \right)
 \ v_{\mu_1} v_{\mu_2} ... v_{\mu_n}\ 
\ \ \ ,
\label{eq:ctlag}
\end{eqnarray}
where
$\xi_\pm = {1\over 2}\left( \xi^\dagger m_q \xi^\dagger\pm \xi m_q^\dagger\xi
\right)$.
In general, the constants $b_i (\mu)$ are renormalization scale dependent,
and must be determined experimentally.
Only the operators with coefficients $b_1$ and $b_3$ contribute to forward
matrix elements in the nucleon.
The operators with coefficients $b_2$ and $b_5$ have at least one
additional pion associated with them, 
while $b_4$ has at least two additional pions associated with it.
The $m_q$-dependence of the matrix elements of 
${\cal O}^{(n), a}_{\mu_1\mu_2\ ... \mu_n}$
shown in eqs.~(\ref{eq:pitotal}), (\ref{eq:Ntotal})
and (\ref{eq:ctlag})
has the form 
\begin{eqnarray}
{\cal M} &\sim & \alpha\ 
+\ \beta m_q\log\left(m_q/\mu\right)\ +\ \gamma(\mu) m_q\ +\ ...
\ \ ,
\end{eqnarray}
where the $\mu$-dependence of 
$\gamma(\mu)$ is precisely equal and opposite that
of the 
$\beta$-term, leaving an expression that is explicitly $\mu$-independent.
While the contributions from the $\beta$-term formally dominate the sub-leading
contributions in the chiral limit when $\mu=\Lambda_\chi$,
there are well-known examples where
such terms are numerically smaller than the sub-leading analytic contributions,
terms analogous to the $\gamma$-term, for the physical values of $m_q$.
The pion-charge radius $\langle r_\pi^2 \rangle$
is such an example, where for the physical values of the
pion mass (and kaon mass in $SU(3)$), the contribution from the 
$\alpha_9 (\mu)$ counterterm (evaluated at $\mu=\Lambda_\chi$) is twice that
of the non-analytic loop contributions of the form $\log m_q$.
Therefore, while the terms we have computed 
in eq.~(\ref{eq:Ntotal}) are the 
formally dominant sub-leading contributions in the 
chiral limit they may not dominate the sub-leading contribution
for physical values of $m_q$ due to the terms shown in 
eq.~(\ref{eq:ctlag}).

\begin{figure}[!ht]
\centerline{{\epsfxsize=3.5in \epsfbox{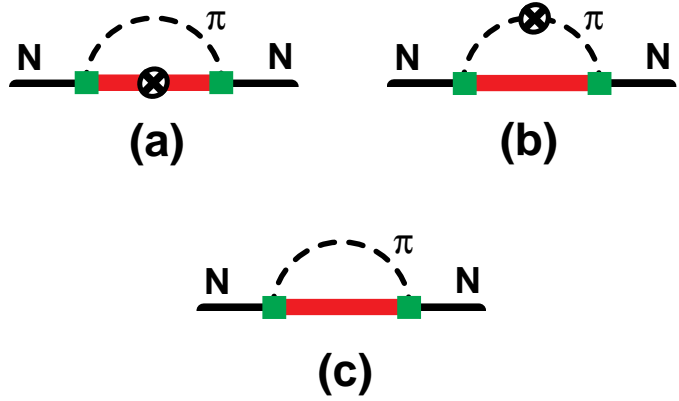}}} 
\vskip 0.15in
\noindent
\caption{\it 
One-loop diagrams with $\Delta$ intermediate states that contribute to the 
matrix elements of ${\cal O}^{(n), a}_{\mu_1\mu_2\ ... \mu_n}$
between single-nucleon states.
The thick solid line inside the loop denotes a $\Delta$ propagator,
while the dashed line denotes  a pion propagator.
The crossed circle denotes an insertion of an operator from
eq.~(\ref{eq:meson}) or eq.~(\ref{eq:deltas}), 
arising directly from the twist-2 operator, and 
the square denotes an insertion of the strong $N\Delta\pi$ 
interaction~$\propto g_{N\Delta}$.
Diagrams (a) and (b) are vertex corrections while diagram (c) denotes
nucleon wavefunction renormalization.
}
\label{fig:deldiags}
\vskip .2in
\end{figure}

At relatively low momentum scales, $\sim 300~{\rm MeV}$, there can be
large contributions from loop diagrams involving the $\Delta$'s.
The formal construction and phenomenology of an EFT with dynamical
$\Delta$'s (or any resonance)
has been studied 
extensively~\cite{JMheavy,chiralN,chiralUlf,chiralMAG,chiralEff,chiralAX,chiralPOL,chiralQ,delother}.
If all diagrams with the 
$\Delta$-resonance as an intermediate state are included
then one can consistently take $\mu\sim \Lambda_\chi$, and 
capture the dominant infrared behavior of the theory~\cite{JMheavy,chiralN}.
Matrix elements of ${\cal O}^{(n), a}_{\mu_1\mu_2\ ... \mu_n}$
between $\Delta$ states are described, at leading order, by
\begin{eqnarray}
{\cal O}^{(n), a}_{\mu_1\mu_2\ ... \mu_n}
\ &\rightarrow &\ 
C^{(n)}\ v_{\mu_1} v_{\mu_2} ... v_{\mu_n}\ 
\overline{\Delta}^\alpha_v \ \tau^a_{\xi +}\ \Delta_{\alpha , v}
\ -\ {\rm traces}
\nonumber\\
&  + &  D^{(n)}\ {1\over n!} v_{\{ \mu_1} v_{\mu_2} ... v_{\mu_{n-2}}\ 
\overline{\Delta}_{\mu_{n-1} , v} \ 
\tau^a_{\xi +}\ 
\Delta_{\mu_n\} , v}
\ -\ {\rm traces}
\ \ \ ,
\label{eq:deltas}
\end{eqnarray}
where $C^{(1)}=-3$ by normalization of the isospin charge operator, 
and $D^{(1)}=0$ simply because of the number of available Lorentz indices.
These operators will contribute to matrix elements 
between nucleon states through loop diagrams.
The leading order 
strong interactions between the $N$'s, $\Delta$'s and $\pi$'s 
are described
by a lagrange density of the form
\begin{eqnarray}
{\cal L}^\Delta & = & 
g_{N\Delta}\ \left[
\overline{\Delta}^{\alpha , ijk}\ {\cal A}_{\alpha , k}^l\  N_j\ \epsilon_{il}
\ +\ {\rm h.c.}\ \right]
\ \ \ ,
\label{eq:stdel}
\end{eqnarray}
where the coupling is $g_{N\Delta}\sim 1.8$, from the observed
width for $\Delta\rightarrow N\pi$~\cite{chiralN}.
The loop diagrams shown in fig.~\ref{fig:deldiags}
gives rise to a forward matrix element between single-nucleon states of 
\begin{eqnarray}
{\cal M} & = & 
-{g_{N\Delta}^2\over 4\pi^2 f^2}\ 
J_1\left(\Delta M, m_\pi\right)\ 
\overline{U}_v \tau^a U_v\ 
\left[\ 
A^{(n)}\ +\ {5\over 9} C^{(n)}\ -\ {5\over 27} D^{(n)} 
\ +\ {2\over 3}\delta^{n 1}
\ \right]
\nonumber\\
& & 
\left(\ v_{\mu_1} v_{\mu_2} ... v_{\mu_n}\ 
\ -\ {\rm traces}\ \right)
\ \ \ ,
\label{eq:deltotal}
\end{eqnarray}
where $\Delta M = M_\Delta - M_N$ is the $\Delta$-N mass difference.
The function $J_1$ is
\begin{eqnarray}
J_1\left(\Delta, m\right)
& = & 
\left( m^2-2\Delta^2\right)\log\left({m^2\over\Lambda_\chi^2}\right)
\ +\ 2\Delta\sqrt{\Delta^2-m^2}
\log\left({\Delta - \sqrt{\Delta^2-m^2+ i \epsilon}
\over \Delta + \sqrt{\Delta^2-m^2+ i \epsilon}}\right)
\ \ \ ,
\label{eq:Jone}
\end{eqnarray}
where we have renormalized at the scale $\mu=\Lambda_\chi$.
In the limit that $\Delta\rightarrow 0$, $J_1$ contains a chiral logarithm,
as is clear from eq.~(\ref{eq:Jone}), while in the limit of 
large $\Delta$ only terms analytic in $m$ survive, as required by the
decoupling of the $\Delta$~\cite{chiralN}.
The reason that such contributions must be included in order to sensibly
renormalize the theory at $\mu=\Lambda_\chi$ is that the scale for the 
$m_q$-dependence from these diagrams is set by  $\Delta M$ and not by 
$\Lambda_\chi$, and therefore a naive estimate of the size of
counterterms in the theory without the explicit $\Delta$-fields is set by 
$\Delta M$ and not $\Lambda_\chi$.
Without resorting to hadronic models, one is unable to make statements
about $C^{(n)}$ or $D^{(n)}$, and they must be determined elsewhere.
It is interesting to note that $N\Delta$ transition operators
induced by  ${\cal O}^{(n), a}_{\mu_1\mu_2\ ... \mu_n}$ involve either a
spin-operator $S^\mu$ or additional pion fields.
Neither type of operator contributes to nucleon matrix elements at 
one-loop level.
Our result in eq.~(\ref{eq:deltotal})
disagrees with the result of Ref.~\cite{CBMa}, as they
computed only the contribution
from  fig.~\ref{fig:deldiags} (b),
where  ${\cal O}^{(n), a}_{\mu_1\mu_2\ ... \mu_n}$ is inserted into the 
pion propagator.


\section{Singlet, Twist-2 Operators}

In this section we consider matrix elements of singlet twist-2 operators,
$^{(S)}{\cal O}^{j (n)}_{\mu_1\mu_2\ ... \mu_n}$ with $j=q,g$ for the 
quark and gluonic operators,
of the form
\begin{eqnarray}
^{(S)}{\cal O}^{q (n)}_{\mu_1\mu_2\ ... \mu_n}
& = & 
{1\over n!}\ 
\overline{q}\ \gamma_{ \{\mu_1  } 
\left(i \stackrel{\leftrightarrow}{D}_{\mu_2}\right)\ 
... 
\left(i \stackrel{\leftrightarrow}{D}_{ \mu_n\} }\right)\ 
q \ -\ {\rm traces}
\nonumber\\
^{(S)}{\cal O}^{g (n)}_{\mu_1\mu_2\ ... \mu_n}
& = & 
{1\over n!}\ 
G^a_{\alpha \{ \mu_1}
\left(i \stackrel{\leftrightarrow}{D}_{\mu_2}\right)\ 
... 
\left(i \stackrel{\leftrightarrow}{D}_{ \mu_{n-1} }\right)\ 
G^{a,\alpha}_{\mu_n\}}
 \ -\ {\rm traces}
\ \ \ .
\label{eq:sings}
\end{eqnarray}
and  it is clear that the $^{(S)}{\cal O}^{j (n)}_{\mu_1\mu_2\ ... \mu_n}$
transform as $(1,1)$ under  $SU(2)_L\otimes SU(2)_R$.

The matrix element of  $^{(S)}{\cal O}^{j (n)}_{\mu_1\mu_2\ ... \mu_n}$
in the pion will be described at leading order by  a lagrange density of the 
form
\begin{eqnarray}
^{(S)}{\cal O}^{j (n)}_{\mu_1\mu_2\ ... \mu_n}
&\rightarrow &
\bar a^{(n)}_j \left(i\right)^n {f^2\over 4}
\left({1\over\Lambda_\chi}\right)^{n-1}
\ {\rm Tr}\left[
\Sigma^\dagger \overrightarrow\partial_{\mu_1}
...\overrightarrow\partial_{\mu_n}
\Sigma
\ +\ 
\Sigma \overrightarrow\partial_{\mu_1}
...\overrightarrow\partial_{\mu_n}
\Sigma^\dagger
\right]
\ -\ {\rm traces}
\nonumber\\
& = & 
\bar a^{(n)}_j 2 \left(i\right)^n
\left({1\over\Lambda_\chi}\right)^{n-1}
\pi^\alpha  
\overrightarrow\partial_{\mu_1}
...\overrightarrow\partial_{\mu_n} \pi^\alpha
\ -\ {\rm traces}
\ +\ {\cal O}\left(\pi^4\right)
\ \ \ ,
\label{eq:singmeson}
\end{eqnarray}
where the $\bar a^{(n)}_j$ are coefficients that must be determined elsewhere,
and depend upon the particular singlet operator under consideration.
A calculation of one-loop diagrams analogous to those shown in 
fig.~\ref{fig:pidiags}
give rise to a matrix element between pions with isospin indices $\alpha$
and $\beta$, at next-to-leading order, of
\begin{eqnarray}
{\cal M}_j & = & 
4\  \bar a^{(n)}_j
\left({1\over\Lambda_\chi}\right)^{n-1}
\ \delta^{\alpha\beta}
\ q_{\mu_1} ... q_{\mu_n}
\ -\ {\rm traces}
\ \ \ ,
\end{eqnarray}
for $n$-even, while the matrix element for $n$-odd vanishes.
There are no non-analytic corrections to this matrix element 
for any $n$ at next-to-leading order.

In contrast to  eq.~(\ref{eq:nucleons}),
the leading-order operator contributing to the matrix element of 
a singlet operator
$^{(S)}{\cal O}^{j (n)}_{\mu_1\mu_2\ ... \mu_n}$
between nucleon states is
\begin{eqnarray}
^{(S)}{\cal O}^{j (n)}_{\mu_1\mu_2\ ... \mu_n}
\ &\rightarrow &\ 
\bar A^{(n)}_j\ v_{\mu_1} v_{\mu_2} ... v_{\mu_n}\ 
\overline{N}_v \ N_v
\ -\ {\rm traces}
\ \ \ ,
\label{eq:singnucleons}
\end{eqnarray}
where operators involving more derivatives or insertions of $m_q$ are 
suppressed by powers of~$\Lambda_\chi$.
Since $^{(S)}{\cal O}^{q (1)}_{\mu}$ is the baryon number operator and 
$^{(S)}{\cal O}^{g (n)}_{\mu_1\mu_2\ ... \mu_n}$ vanishes for $n<2$, we have
that $ \bar A^{(1)}_q=+3$ and $\bar A^{(1)}_g=0$.
Only some of the one-loop
diagrams in fig.~\ref{fig:Ndiags} contribute to matrix
elements of the singlet operators.
Diagrams (c) and (d)  of fig.~\ref{fig:Ndiags} are absent while
diagram (b) can only contribute at higher orders.
Further, the contribution from the vertex diagram, diagram (a), is exactly
canceled by the contribution from wavefunction renormalization, diagram (e).
Therefore, the singlet matrix elements in the nucleon 
do not receive any non-analytic
corrections of the form $m_q\log m_q$ from nucleon intermediate 
states~\footnote{The loop corrections are the
same as those contributing to the nucleon mass, for which there are no terms
of the form  $m_q\log m_q$.}.
However, there are contributions from $\Delta$ intermediate states.
The leading order matrix elements involving the $\Delta$ are described by
the operators
\begin{eqnarray}
^{(S)}{\cal O}^{j (n)}_{\mu_1\mu_2\ ... \mu_n}
\ &\rightarrow &\ 
\bar C^{(n)}_j\ v_{\mu_1} v_{\mu_2} ... v_{\mu_n}\ 
\overline{\Delta}^\alpha_v \ \Delta_{\alpha , v}
\ -\ {\rm traces}
\nonumber\\
&  + &  \bar D^{(n)}_j\ {1\over n!} v_{\{ \mu_1} v_{\mu_2} ... v_{\mu_{n-2}}\ 
\overline{\Delta}_{\mu_{n-1} , v} \ \Delta_{\mu_n\} , v}
\ -\ {\rm traces}
\ \ \ ,
\label{eq:singdeltas}
\end{eqnarray}
with 
$ \bar C^{(1)}_q=-3$, $\bar C^{(1)}_g=0$,
$ \bar D^{(1)}_q=0$, and $\bar D^{(1)}_g=0$.
These operators contribute through loop-diagrams
shown in fig.~\ref{fig:deldiags}.
Diagram (b) of  fig.~\ref{fig:deldiags} contributes only at higher orders
in the chiral expansion, and we find a contribution to the nucleon
matrix element of
\begin{eqnarray}
{\cal M}_j & = & 
-{g_{N\Delta}^2\over 4\pi^2 f^2}\ 
J_1\left(\Delta M, m_\pi\right)\ 
\overline{U}_v U_v\ 
\left[\ 
\bar A^{(n)}_j\ +\ \bar C^{(n)}_j\ -\ {1\over 3}\bar D^{(n)}_j 
\ \right]
\nonumber\\
& & 
\left(\ v_{\mu_1} v_{\mu_2} ... v_{\mu_n}\ 
\ -\ {\rm traces}\ \right)
\ \ \ .
\label{eq:singdeltotal}
\end{eqnarray}
As required, corrections to the 
matrix element of the baryon number operator,
$^{(S)}{\cal O}^{q (1)}_{\mu}$, vanish.

\section{Conclusions}

We have computed the leading non-analytic contributions of the form 
$m_q\log m_q$
to matrix elements of twist-2 operators that arise 
in deep-inelastic scattering.
A previously omitted contribution to the matrix elements of non-singlet 
operators that is independent of $g_A$ and 
related by chiral symmetry to the tree-level vertex is identified.
Our results will aid in the extrapolation of 
unquenched lattice calculations 
of single-nucleon matrix elements of twist-2 operators from the
quark masses used on the lattice to their physical values.

The work we have presented here can be straightforwardly extended to 
off-forward matrix elements of the twist-2 operators,
i.e. those in which there is a non-zero momentum transfer to the hadronic 
system from the twist-2 operator.
Such matrix elements have received significant amount of attention during the 
past few years, as one can define off-forward parton distributions 
as a simple extension of the parton model 
(for an overview see Ref.~\cite{ji}).
In addition, deeply-virtual-Compton-scattering (DVCS) has been extensively
explored as a means to measure such distributions.
The effective field theory construction we have employed in this work will
allow for a description of these matrix elements in the low-momentum regime.

\vskip 0.5in

We thank Jiunn-Wei Chen for useful discussions.
This work is supported in
part by the U.S. Dept. of Energy under Grant No.  DE-FG03-97ER4014.

\end{document}